\documentclass[twocolumn,showpacs,superscriptaddress,amsmath,amssymb,aps,pra]{revtex4-1}
\usepackage{graphicx}
\usepackage{CJKutf8}
\usepackage{dcolumn}
\usepackage{amsmath,bm}
\usepackage{epstopdf}
\usepackage[mathlines]{lineno}
\usepackage{color}
\usepackage[colorlinks=true,linkcolor=blue,citecolor=magenta,urlcolor=cyan]{hyperref}

\begin{document}
\title{Fractional Charges in the Su-Schrieffer-Heeger Model}
\author{Yi-Dong Wu}
\email{wuyidong@ysu.edu.cn}
\affiliation{Key Laboratory for Microstructural Material Physics of Hebei Province, School of Science, Yanshan University, Qinhuangdao 066004, China}

\begin{abstract}
The Su-Schrieffer-Heeger(SSH) model has been widely used to study the topological property of 1D systems. It is claimed that there is fractional charge at the boundary of the nontrivial phase while none at that of trivial phase. However, this conclusion is in direct contradiction to the modern theory of polarization(MTP). We solve this paradox by showing that the polarization of SSH model depends only on the distribution of the positive charges and is irrelevant to the Zak phase defined in previous works. Thus the distribution of positive charges alone determines whether the SSH chain is topological or not. Similarly, we show the polarization defined by Berry connection can not be used to characterize topological property of a 2D generalization of SSH model.
\end{abstract}
\date{\today}
\maketitle
Electric polarization is a fundamental concept in condensed matter physics. However, it has long been controversial that whether the electric polarization in solids is a well-defined bulk property. The ambiguities in defining the electronic polarization are removed after King-Smith and Vanderbilt  proposed the MTP\cite{PhysRevB.47.1651,PhysRevB.48.4442,RevModPhys.66.899}. In this theory the electronic polarization is defined in terms of the Berry phase of the occupied energy bands, or equivalently in terms of the Wannier centers. With the MTP the polarization itself becomes well-defined as a bulk property and its relation to the surface charge is established\cite{PhysRevB.48.4442}.

In a 1D system we have $\sigma=\pm P$(modulo $e$), where
$\sigma$ is the charge at the boundary and $P$ is the bulk polarization. If $P$ takes a fractional value in unit of $e$ there will be fractional charge at the boundary of the system. Thus it is suitable to use MTP to study an interesting phenomenon in condensed matter system---charge fractionalization.

It has been proposed that a  topological soliton excitation located between two phases of the SSH model may carry a fractional charge $q=\pm e/2$ if the electrons are spinless\cite{PhysRevLett.42.1698,PhysRevLett.46.738}. Later it is claimed that the two phases of the SSH model are topologically distinct. The Zak phase $\phi_{Zak}$ is used to characterize the topological property of the two phases of the SSH model\cite{PhysRevLett.89.077002,PhysRevB.84.195452, Poli2015,atala2013direct,PhysRevLett.102.065703,PhysRevLett.115.040402,PhysRevLett.110.180403}. The phase with $\phi_{Zak}=\pi$ and zero-energy edge states at its ends is considered as a 1D topological insulator while the phase with $\phi_{Zak}=0$ is considered topologically trivial. It is also claimed that the polarization $P$ is related to Zak phase by $P=-e\phi_{Zak}/2\pi$\cite{PhysRevB.96.245115}. Then $P$ can be considered as a topological invariant, which can be used to predict whether there is fractional charge at the boundary of a SSH chain. The polarization defined by Berry connection has also been used as a topological invariant to characterize a 2D generalization of SSH model, which is claimed to be a high order topological insulator\cite{PhysRevLett.120.026801}.

One important consequence of the MTP is that the bulk polarization(modulo $e$ in 1D case) does not depend on the choices of the unit cell of the bulk system. It is because the Wannier center is independent of the choice of unit cell and a change in the choice of unit cell only leads to a translation of an integer positive charge by a lattice vector. Another merit of the MTP is that in 1D system the boundary charge determined by the polarization is independent of how the system is terminated as long as the edge states are either full or empty if they exist.

In this letter we show the conclusion that there are fractional charges at the edge of the nontrivial phase of the SSH chain while none at that of the trivial phase is in direct contradiction to the MTP. The contradiction comes from the definition of the polarization in the SSH model. First, the Zak phase defined in Ref.\cite{PhysRevLett.89.077002,PhysRevB.84.195452, Poli2015,atala2013direct,PhysRevLett.102.065703,PhysRevLett.115.040402,PhysRevLett.110.180403} does not correspond to the actual Wannier center. Second, contribution from the positive charges is not specified in discussing the polarization of the SSH model.

To study the charge at the boundary by using polarization, the bulk system must be charge-neutral. However, if the electrons are spinless, to make the bulk system charge-neutral  the positive charges in one unit cell must be $e$ on average when the Fermi level lies in the bulk gap. Thus we can no longer assume each ion carry a charge $e$.  Additional assumptions on the distribution of the positive charges must be made to discuss the polarization of the spinless SSH model. We show that it is the positive charge distribution instead of the electronic property that determines the charge at the boundary of SSH model.

To illustrate this point of view we consider the SSH chain in Fig.1(a). For a given choice of unit cell it is claimed that the system is nontrivial when the magnitude of the intercell hopping is greater than that of the intracell hopping. Otherwise, the system is trivial. The Zak phase $\phi_{Zak}=\pi$ and the polarization $P=e/2$ in the nontrivial phase while $\phi_{Zak}=0$ and  $P=0$ in the trivial phase.

\begin{figure}
  \includegraphics[width=9cm]{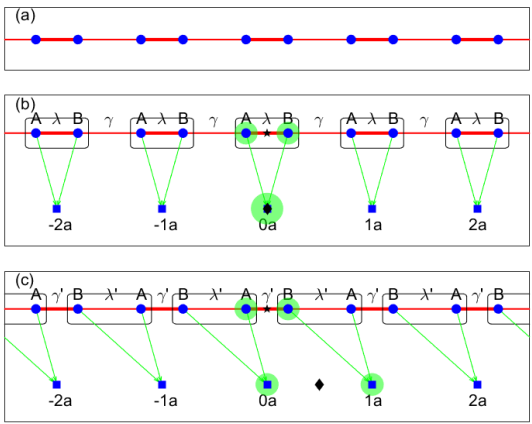}
    \caption{Same bulk SSH model with different choices of unit cells. (b) and (c) Two different choices of unit cells of the same bulk SSH model in (a). Sites in the same rectangle are in the same unit cell. The probability distribution of the Wannier function is represented by the areas of the circle surrounding the sites.  Probability distributions of the Wannier function if all the sites in the same unit cell are assumed to be located at the lattice point are illustrated in (b) and (c). The black pentagrams show the actual positions of Wannier center. The black diamonds show the Wannier centers calculated from the Zak phase. Here $\lambda=\gamma'$, $\gamma=\lambda'$. In this paper we assume $|\lambda|>|\gamma|$.}
    \label{fig1}
 \end{figure}

However, as shown in in Fig.1(b) and 1(c), there are two equally possible choices of unit cells for the SSH chain in Fig.1(a).
With the two choices of unit cells the relative magnitudes between the intercell and intracell hopping are different. Thus the system will be considered as trivial with the choice in Fig.1(b) and nontrivial with choice in Fig.1(c). The polarizations  will differ by $e/2$ with the two choices of unit cells if we use $\phi_{Zak}$ to calculate polariztion.  On the other hand, according to the MTP, the polarizations calculated with different choices of unit cell can only differ by a multiple of $e$. So here is an apparent contradiction between the two widely accepted theories.

In the MTP the contribution of the electronic charges to the polarization equivalents to that of localized point charges $-e$ located at the Wannier centers associated with the occupied bands in each unit cell. If there is only one occupied band as in the spinless SSH model, the Wannier center is an invariant(modulo lattice constant $a$), which does not depend on the choice of the unit cell\cite{PhysRevB.47.1651,PhysRevB.48.4442,RevModPhys.66.899}. However, if we use the Zak phase $\phi_{Zak}$\cite{PhysRevB.96.245115} to calculate  the Wannier center $\tilde{x}_w$  by $\tilde{x}_w=a\phi_{Zak}/2\pi$ the Wannier centers will differ by $a/2$ with the two choices of unit cells.

To find out which theory gives the correct prediction we construct the Wannier function of the SSH model explicitly. Though there are infinite ways to construct the Wannier function, the Wannier center is invariant modulo a lattice constant $a$. Thus one choice of Wannier function is enough for our discussion. Here we choose a Wannier function that makes the probability distribution symmetrical about the midpoint of the bond with greater hopping amplitude. Thus the Wannier center is always located at the midpoint of the strong bond. Then the coordinate of the Wannier center $x_w$ depends only on the location of origin or the lattice point $R=na$\cite{1706.02370}.

Now we explain why the Wannier center $\tilde{x}_w$ derived from $\phi_{Zak}$ can take two values $0$ and $a/2$(modulo $a$) with different choices of unit cells. We denote the Wannier function of the occupied band as $W_i(n)$, where $n$ denotes the $n$th unit cell and $i=A$ and $B$ indicates the two sites within a unit cell as shown in Fig.1(b) and 1(c). It can be shown that $\tilde{x}_w$ can be expressed as
\begin{equation}\label{zak}
\begin{split}
\tilde{x}_w &=a\frac{1}{2\pi}i \int_0^{2\pi}\langle u (k)|\frac{\partial}{\partial k}|u (k)\rangle dk\\
             &=\sum\limits_{n}na (|W_A(n)|^2+|W_B(n)|^2),
\end{split}
\end{equation}
where $|u(k)\rangle$ is the normalized eigenstate of Bloch Hamiltonian of SSH model in momentum space. It can be seen from this expression for $\tilde{x}_w$ that the two sites in one unit cell are assumed to be both located at the lattice point $R=na$ in calculating the average position.

If the Wannier function have the expression $W_i(n)$ with the unit cell in Fig.1(b) it will have an expression $W'_i(n)$ which satisfies $W'_B(n)=W_B(n-1)$ and $W'_A(n)=W_A(n)$ with the choice of unit cell in Fig.1(c). Then the ``Wannier center'' will have the expression
\begin{equation}\label{zak}
\begin{split}
\tilde{x}'_w &=\sum\limits_{n}na (|W'_A(n)|^2+|W'_B(n)|^2)\\
             &=\sum\limits_{n}na |W_A(n)|^2+(n+1)a|W_B(n)|^2.
\end{split}
\end{equation}
Thus with the two choices of unit cells the difference of the ``Wannier center'' $\tilde{x}'_w-\tilde{x}_w=a\sum_n |W_B(n)|^2$. It is easy to show that $\sum_n |W_B(n)|^2=1/2$ . So the difference between the ``Wannier centers'' with the two choices of unit cell is $a/2$.

It is clear the ``Wannier centers'' $\tilde{x}_w$ does not correspond to the actual average position of the electron described by a Wannier function. The two values of $\tilde{x}_w$ or $\phi_{Zak}$ only reflect the two choices of unit cells of the same bulk system. If we insist that the finite system contain integer number of the unit cells the Zak phases may be related to the existence of edge states\cite{PhysRevLett.89.077002,PhysRevB.84.195452}. However, the $\tilde{x}_w$ or $\phi_{Zak}$ can not be used to study the polarization of the SSH chain.

Now the question is: whether there are two topologically distinct phases of the SSH model if we only consider the electrons. For the bulk system the answer is simple: no, because if the answer is yes different observers studying the same same bulk system will get opposite conclusions on the topology of the system based on the unit cell they choose. So there is only one bulk phase of the electrons of the SSH model. In fact a similar view has been expressed in Ref.\cite{PhysRevLett.46.738} by stating ``B(phase) is simply a translation of A(phase) by one lattice spacing($a/2$)''. The experimental realization of the SSH model confirm this point of view. The lattice potentials for topologically ``trivial'' and ``nontrivial'' phases in Ref.\cite{atala2013direct} differ only by a translation or a different choice of origin of coordinate.

For finite system one may prefer to consider the two systems in Fig.2(b) and 2(c) as topologically distinct based on the existence or absence of edge states. However, this classification may lead to confusion in some cases. For example, there is edge state at one end of the system in Fig.2(d) while none at the other end. Then it is not clear whether this system is nontrivial or not. Thus it is more convenient to consider the systems in Fig.2 as the same bulk phase with different choices of terminations.
 \begin{figure}
  \includegraphics[width=9cm]{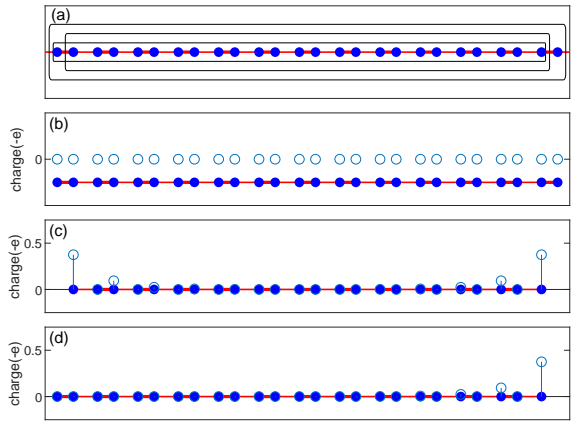}
    \caption{(b)-(d)Different choices of boundaries of the bulk SSH chain in (a). The net charge at each site is illustrated in (b)-(d). Each site is assumed to carry a positive charge $e/2$. The Fermi level lies in the bulk gap and the edge states are full in (c) and (d). Here $\lambda=-2$ and $\gamma=-1$. }
    \label{fig2}
 \end{figure}

It is also claimed that the existence of the factional boundary charge $\pm e/2$ in a SSH chain is an indicator that the system is topologically nontrivial. For example, it is claimed that there is factional charges at the system in Fig.2(c) while none at the boundary of the system in Fig.2(b). However, if we consider the two systems in Fig.2(b) and 2(c) and as the same bulk system with different choice of terminations the boundary charges of the two system can only differ by a multiple of $e$ based the MTP. Thus here is another discrepancy between the two theories.

This mystery can only be solved when the distribution of positive charges are specified. To maintain the bulk charge-neutrality each site must carry $e/2$ if we assume the positive charges are distributed evenly among the sites. If we make this unrealistic assumption, the polarization of the system is only well defined modulo $e/2$. This is because a change in choice of the unit cell induces a translation of positive charge in multiples of $e/2$ while leaving the Wannier center of the electron unaffected. So the polarization may differ by $e/2$ with different choices of unit cells in Fig.1. By the same argument in Ref.\cite{PhysRevB.48.4442} the charge at the boundary is determined modulo $e/2$. That is, the boundary charge may differ by a multiple of $e/2$ when terminated differently. The net charge distributions in Fig.2(b) and 2(c) show there is fractional charge at the boundary of ``nontrivial'' phase while none at that of the ``trivial'' phase. However, it is clear this difference comes solely from the unrealistic assumption about the positive charges instead of from the difference of the Zak phases of the electrons.

 \begin{figure}
  \includegraphics[width=9cm]{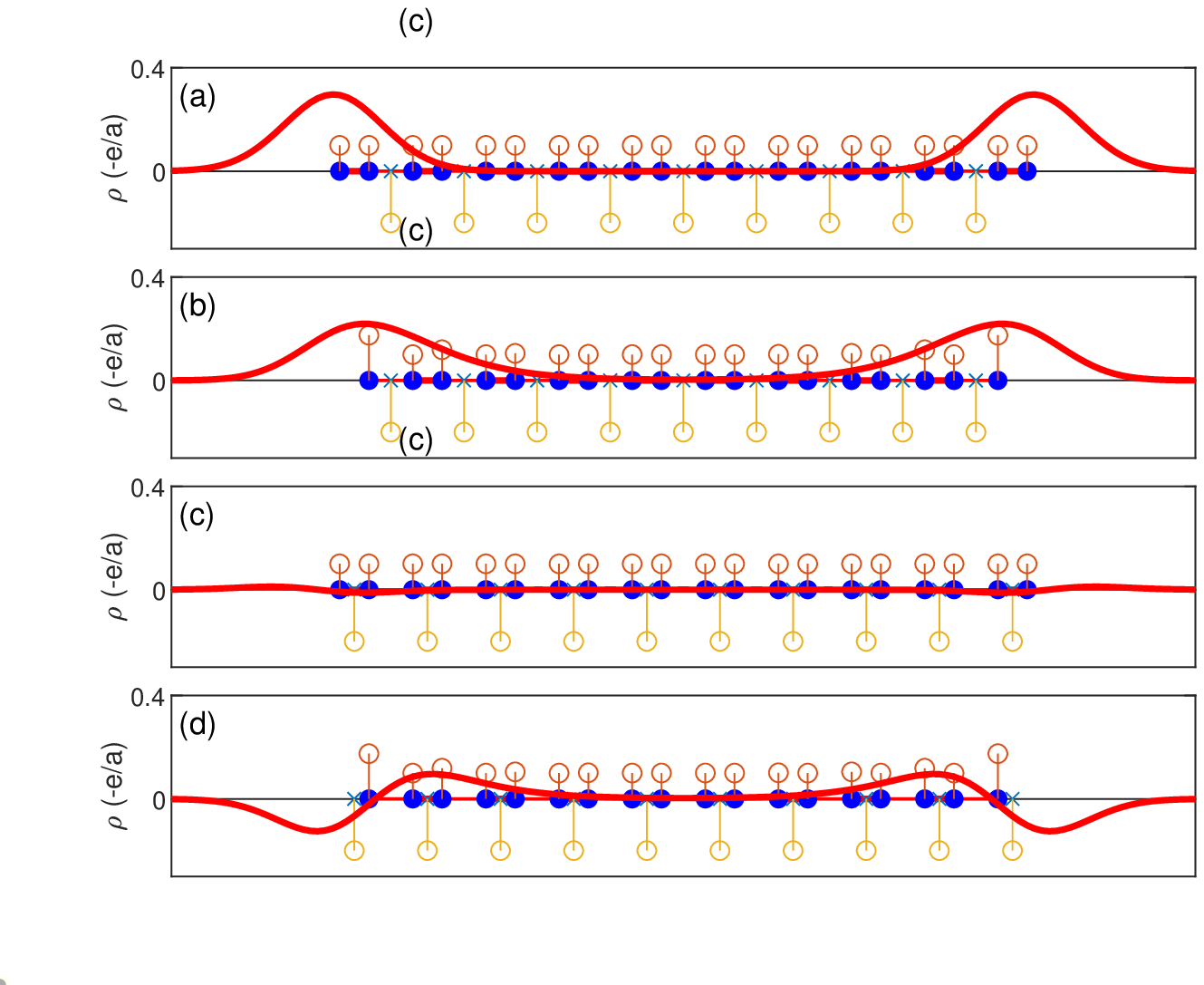}
    \caption{Slow-varying ``macro'' charge density $\rho$ of the SSH chains. The positive point charges are represented with crosses. In (a) and (b) the positive charges is located at the midpoints of weak bonds while in (c) and (d) at midpoints of strong bonds. The edge states in (b) and (d) are full. Here $\lambda=-2$, $\gamma=-1$ and $D=a$. }
    \label{fig3}
 \end{figure}

If we assume that there is one point positive charge $e$ in each unit cell the position of the positive charge becomes important. For example, if we require the SSH model preserves the inversion symmetry there are only two choices of the locations of the positive charges. In this case the MTP works. If the positive charges are located at the midpoints of the weak(strong) bonds as shown in Fig.3 the polarization $P$ of the will be $e/2$($0$) modulo $e$. In this case $P$ can be considered as a topological invariant because it is independent of the choices of unit cells. The SSH chain is a genuine 1D topological insulator when $P=e/2$. The topological property is protected by the inversion symmetry. That is the nontrivial phase can not be continuously connected to the trivial phase without breaking the inversion symmetry.

To study the charge at the boundary we filter out the fast-varying part of the charge density by convolute it with $\frac{\sqrt{\pi}}{D}e^{-x^2/D^2}$\cite{jackson2012classical}.  After this process we get a slow-varying ``macro'' charge density $\rho$. The charge at the boundary can be obtained by integrating $\rho$. In the nontrivial phase($P=e/2$) there are $e/2$ at the boundary regardless the existence of the edge states or not as shown in Fig.3(a) and 3(b). In comparison, the charge at the boundary of the trivial phase($P=0$) in Fig.3(c) and 3(d) is zero. This conclusion about the charge at boundary does not change when more complicated boundary conditions are used\cite{Supplemental}. It can be shown that the fractional charges are robust against random disorders\cite{Supplemental}.

 \begin{figure}
  \includegraphics[width=9cm]{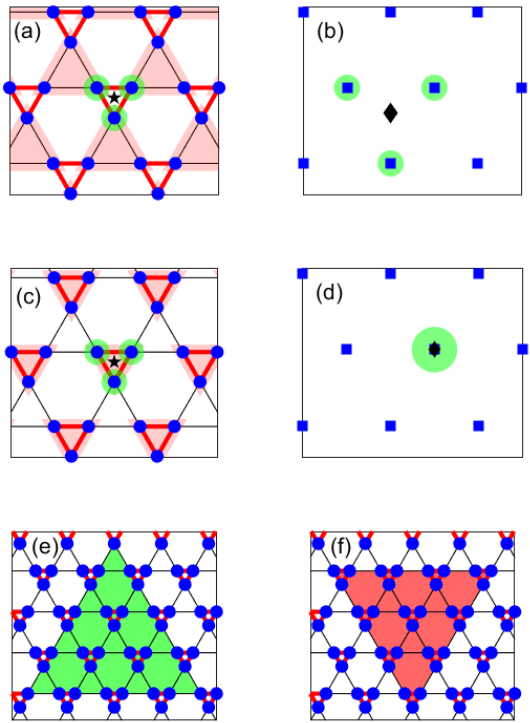}
    \caption{(a) and (c) Same bulk breathing kagome lattice with different choices of unit cells. Orbits in the same red triangle are in the same unit cell. The probability distribution of the Wannier function is represented by the areas of the circle surrounding the orbits. (b) and (d) illustrate the probability distribution of the Wannier function if all the orbits in the same unit cell are assumed to be located at the lattice point. The black pentagrams show the actual position of Wannier center. The black diamonds show the Wannier centers calculated from the Berry connection. (e) and (f)two choices of boundaries of the same bulk system. }
    \label{fig4}
 \end{figure}

As a generalization of the SSH model a 2D model in a breathing kagome lattice is proposed in Ref.\cite{PhysRevLett.120.026801}.  The polarization or Wannier center of the occupied band is used to distinguish two topological phases of the system.  It is claimed that the Wannier center is located at the the lattice site in the trivial phase and at the center of a triangle in the nontrivial phase. However, by constructing the Wannier function we show the actual Wannier center is always located at the center of the triangle formed by the three adjacent strong bonds as shown in Fig.4(a) and 4(c).

Similar to the SSH model the two phases are only the same bulk system with different choices of unit cells as shown in Fig.4(a) and 4(c). The difference of Wannier centers between the two phases also comes from the definition of Wannier center as in the SSH model. By using the definition of Wannier center in Ref.\cite{PhysRevLett.120.026801} the three sites in one unit cell will be considered as being located at the same lattice point as shown in Fig.4(b) and 4(d). Thus the difference of polarizations between the two phases also only reflects two choices of unit cells. It is clear there is no topological distinctions between the two phases if we only consider the bulk system.

One may consider the finite systems as topologically distinct based on the existence of corner states or not. However it more convenient to consider them as the same bulk system with different choices of boundaries as shown in Fig.4(e) and 4(f). It is also claimed that there are fractional charges at the corners of the nontrivial system. However, the bulk charge-neutrality and the role of the positive charges are not discussed. As in the SSH model, the charge at boundary can not be determined if the distribution of the positive charges is not specified.

Similar problems also exist in the high order topological insulator that is not characterized by the polarization\cite{Benalcazar61}. In this case we must also check whether the topological invariant depends on the choices of unit cells and  specify the positive charge distribution in calculating the charge at the corner.

\bibliographystyle{apsrev4-1}
\end{document}